\def\year{2020}\relax
\documentclass[letterpaper]{article} 
\usepackage{aaai20}  
\usepackage{times}  
\usepackage{helvet} 
\usepackage{courier}  
\usepackage[hyphens]{url}  
\usepackage{graphicx} 
\usepackage{natbib}  
\usepackage{caption} 
\usepackage{amsmath,amssymb,amsfonts} 
\usepackage{chngpage} 
\usepackage{booktabs} 
\usepackage{multirow}
\title{Explainable Student Performance Prediction With Personalized Attention for Explaining Why A Student Fails}
\author {
    Kun Niu,
    Xipeng Cao,
    Yicong Yu\\
    School of Computer Science (National Pilot Software Engineering School),\\ Beijing University of Posts and Telecommunications, Beijing\\
    \{niukun, xpcao, yicongyu1012\}@bupt.edu.cn
}

\begin{document}

\maketitle

\begin{abstract}
    As student failure rates continue to increase in higher education, predicting student performance in the following semester has become a significant demand. Personalized student performance prediction helps educators gain a comprehensive view of student status and effectively intervene in advance. However, existing works scarcely consider the explainability of student performance prediction, which educators are most concerned about. 

    In this paper, we propose a novel Explainable Student performance prediction method with Personalized Attention (ESPA) by utilizing relationships in student profiles and prior knowledge of related courses. The designed Bidirectional Long Short-Term Memory (BiLSTM) architecture extracts the semantic information in the paths with specific patterns. As for leveraging similar paths' internal relations,  a local and global-level attention mechanism is proposed to distinguish the influence of different students or courses for making predictions. Hence, valid reasoning on paths can be applied to predict the performance of students. The ESPA consistently outperforms the other state-of-the-art models for student performance prediction, and the results are intuitively explainable. This work can help educators better understand the different impacts of behavior on students' study.
\end{abstract}

\section{Introduction}
The higher education environment is more liberal than others, leaving students with high rates of failure. An enduring issue in higher education is to accurately predict students' performance after tracking their learning and behavior data \cite{Spector18}. One significant application of student performance prediction is to allow educators to monitor students' learning status. Consequently, educators could identify at-risk students to provide timely interventions and guide them through their studies to graduate \cite{XingGPG15}.

With the rapid growth of educational data, processing massive amounts of data requires more complex algorithmic sets and data process methods \cite{DuttIH17}. Prior works generally focus on Massive Open Online Courses (MOOCs) interaction data during a short period \cite{AbdelrahmanW19,Yeung19,AiCGZWFW19,VieK19,LiuYCSZY20}, while real-world teaching scenarios tend to have higher teaching quality and more extended periods. Nowadays, educational data has become more heterogeneous with multiple sources, and a large amount of student interaction data has been retrieved. By observing the data, we noticed that students' semester performance changes dynamically and is prominently affected by their behavior. It is essential to utilize students' short-term behavioral preferences during the course and long-term behavioral habits since students enroll. This forces researchers to build longer and deeper sequential models. In current works, educators only get black box-like performance predictions that are unconvinced. Therefore, it is significant to show the model's prediction basis and explain which behaviors principally affect the students' performance.


\begin{figure}[]
    \centering
    \includegraphics[width=\linewidth]{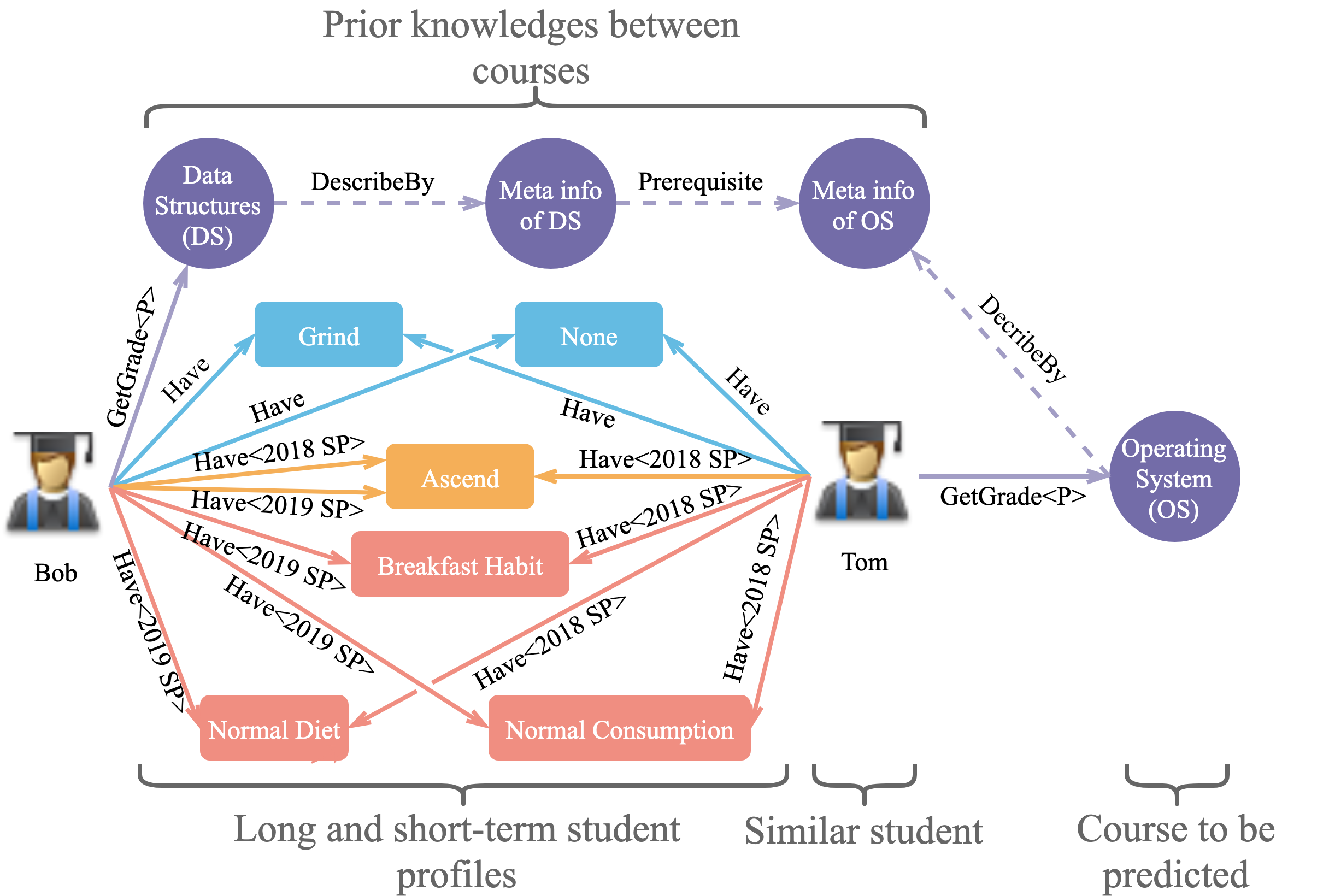}
    \caption{Illustration a sketch of the long short-term (all semesters and each semester) student profiles reflects the relationship between students, students-courses, and courses. The dashed line indicates a prior knowledge between courses, and the solid lines denote the relationships between students and courses.}
    \label{student_profiles}
\end{figure}

Moreover, there are significant similarities between students and courses, e.g., the phenomenon of birds of a feather flock together is common in high education. Furthermore, university courses are often related and have certain similarities. It has proved that the student performance prediction task is quite similar to recommendation problems \cite{Bydzovska15,SuLLHYCDWH18}. We leverage the idea of collaborative filtering \cite{SweeneyLRJ16} in recommender systems to predict performance through students with similar behavior. 

To discover such similarities and solve the problem mentioned above of difficulty modeling long short-term data, we explicitly construct the connections between students and courses using the long short-term student profile and the course knowledge graph. The student profile is calculated from the students' long-term behavior habits (including the students' learning and living status since enrollment) and short-term behavior preferences (including the students' learning and living status in a specific semester). The course knowledge graph contains the prior knowledge between the courses and the courses' meta-information, e.g., Data Structure (DS) is a prerequisite course for Operating System (OS).

For instance, figure \ref{student_profiles} indicates that Bob and Tom have several same tags in the student profile. They have similar habits and similar academic records. Besides, Bob and Tom belong to different grades of the same major, and Tom is higher than Bob. Cause Tom passed the OS and Bob passed the DS, we can infer that Bob may also pass the OS. Conversely, if the model predicts that Bob will pass the OS, we wish to know why the model makes decisions. 

To fill the gap in the lack of explainability of current performance prediction approaches, we propose a novel Explainable Student performance prediction method with Personalized Attention (ESPA). The heart of our work is a path encoder and a personalized attention module. In the path encoder, we use several Bidirectional Long Short-Term Memory (BiLSTM) networks to learn the representations of student-course paths in the student profile and course knowledge graph. Since different students, courses, and behaviors may have different informativeness for prediction. Whereafter, we notice that even the same behavior (such as late sleeping) may affect their studies for different students. A local- and global-level attention mechanism is designed to distinguish these differences. Extensive experimental results on a real-world dataset validate the effectiveness of our approach on student performance prediction.

Our main contributions are listed below:
\begin{itemize}
    \item We propose a novel hierarchical attention mechanism for student performance prediction, combined with knowledge graphs to provide significant explainability.
    \item We provide a solution to complete student performance prediction in an end-to-end manner with a large amount of heterogeneous student interaction data. Our model has the potential to be extended to other tasks.
    \item Experimental results on a real-world dataset show that ESPA outperforms most state-of-the-art approaches. It is also highly explainable for explaining why the model predicts one student may fail in the examination.
\end{itemize}

\begin{figure*}[h]
    \centering
    \includegraphics[width=0.95\linewidth]{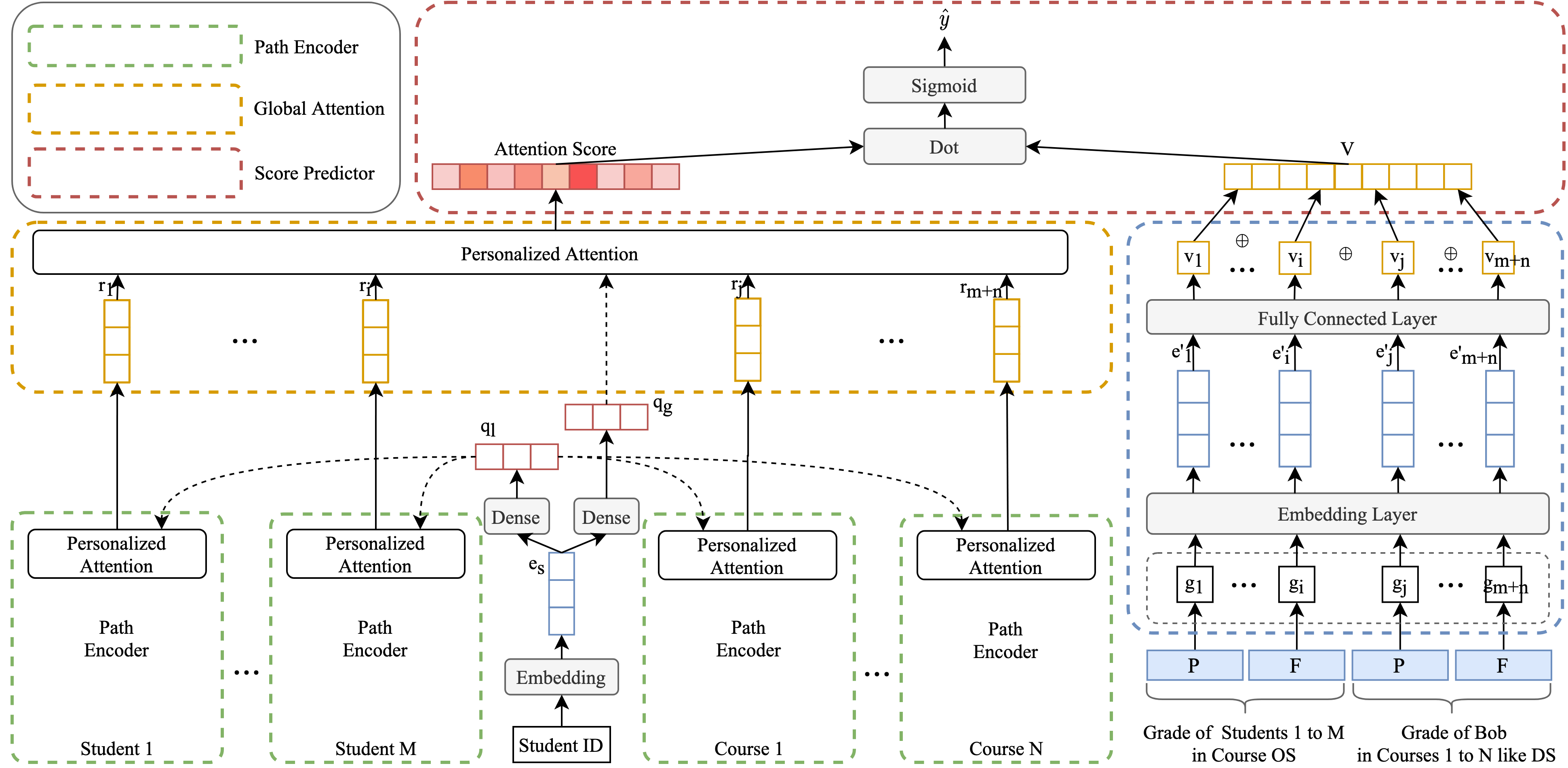}
    \caption{The framework of our approach for explainable student performance prediction. The Path Encoder part will be detailed in the third subsection. The query of hierarchical attention mechanism is the embedding of student id. We leverage student id to extract the most related paths of each scale. Finally, the attention scores to weight the performance representations of similar students to predict the student whether fails.}
    \label{model}
  \end{figure*}

\section{Related Works}
\subsection{Performance Prediction}
\paragraph{Recommender-based methods} \cite{Thai-NgheDKS10} proposed using matrix factorization and collaborative filtering techniques in recommender systems for predicting student performance. The work of \cite{Bydzovska15} applies collaborative filtering to student performance prediction. Furthermore, \cite{SweeneyLRJ16} used a recommender method to predict the next term student performance and find the collaborative filtering based method to achieve the lowest prediction error. Similarly, \cite{HeLZNHC17} proposed the neural architecture for binary classification or recommendation tasks with implicit feedback. These works proved the feasibility of the recommender-based method, and the similarity between students can be utilized to predict student performance.

\paragraph{Deep learning methods} Recent researches focused on leverage deep learning methods to improve prediction performance, \cite{KimVG18, KimVG182} recasted the student performance prediction problem as a sequential event and proposed CritNet for extracting student learning status during a course. \cite{SuLLHYCDWH18,LiuHYCXSH19} proposed an Exercise-Enhanced Recurrent Neural Network framework with an attention mechanism, which leverages the sequence of exercises for inferring scores. This work inspired us to leverage the representations of the student profile for predicting student performance. In recent years, there are several deep learning methods using data such as knowledge graphs\cite{HuangZDWC18,WangWX00C19,XianFMMZ19,XianFZGCHG0MMZ20}, comments\cite{LuoSMG15,DascaluPBCT16} as side information for providing explainability. Inspired by these works, we utilize the long short-term student profile and the relationships between the courses, which have been generally ignored in the educational field.

\subsection{Path-based Methods}
In the literature on path-based methods, \cite{ZhouHHZT18} first clustered a collection of learners and trains the LSTM model to predict their learning paths and performance. As for the knowledge graph, \cite{CatherineMEC17} proposed a method using a knowledge graph to produce recommendations and predictions with the explanations. Moreover, \cite{WangWX00C19} contributed a novel model named Knowledge-aware Path Recurrent Network (KPRN) to utilize a knowledge graph for the recommendation. Inspired by KPRN, we generate representations for paths by accounting for both entities and relations and perform reasoning based on paths. However, KPRN does not take into account the connection between users and the individualized preference of users. Furthermore, our approach mainly considers the similarity between students and courses. ESPA can perform personalized performance prediction with the hierarchical attention mechanism. At present, for short meta-paths with specific patterns, the Recurrent Neural Network (RNN)-based approach is still efficient. Following the previous works, we still leverage the RNN-based method to extract the representations of the paths in the student profile and the knowledge graph of courses.

\section{Method}
In this section, we elaborate on our approach to student performance prediction. We translate student performance predictions into a binary classification problem (positive sample 1 represents failure, negative sample 0 means pass a course) since educators are more concerned with students at risk of failing. Before introducing the model, we first define student-course paths and prior knowledge of courses formally.

\subsection{Student paths}
A knowledge graph is a directed graph composed of entities and relationships. In the course knowledge graph, we use $\mathcal{C}=\{c_i\}^{C'}_{i=1}$ ($C'$ is the number of courses) to denote the set of courses and its meta-information. And $\mathcal{R}_C=\{r_i\}^{R'}_{i=1}$ to represent relationships between courses. The knowledge graph is generally represented as triples of nodes and edges such as $\{(c_1,r,c_2)|c_1,c_2\in\mathcal{C},r\in\mathcal{R_C}\}$, where entities $c_1,c_2$ indicates the start node and end node. And $r$ represents the relationship between two courses (e.g., course DS is a required course for the OS).

For student profile, $\mathcal{S}=\{s_i\}^{S'}_{i=1}$ and $\mathcal{T}=\{t_i\}^{T'}_{i=1}$ sepa-\\rately denote the student set and the tag set in the student profile. The tag set contains student dynamic learning status and behavioral preferences in the student profile. We also define a relationship set $\mathcal{R}_S=\{have,belong\_to,get\_grade,in\}$. Following the work of \cite{ChaudhariAM17}, we define student-tag relationships in student profile with $\{(s,have,t)|s\in\mathcal{S},t\in\mathcal{T}\}$ and $\{(t,belong\_to,s)|t\in\mathcal{T},s\in\mathcal{S}\}$. Furthermore, student-course relationships are defined as $\{(s,get\_grade,grade)|s\in\mathcal{S},score\in\{P(pass),F(fail)\}\}$ and $\{(grade,in,c)|s\in\mathcal{S},c\in\mathcal{C}\}$.

We merge the course knowledge graph, the student profile, and student-course relationships as a final point. Thus we get a comprehensive knowledge graph $\mathcal{KG}=\{(e_1,r,e_2)|e_1,e_2\in\mathcal{E},r\in\mathcal{R}\}$ where $\mathcal{E}=\mathcal{C}\cup\mathcal{S}\cup\mathcal{T}\cup\{score\}$ and $\mathcal{R}=\mathcal{R_C}\cup{R_S}$. For consistency, the knowledge graph $\mathcal{KG}$ in the rest paper denotes the combined graph.

\subsection{Performance inference from similar students and prior courses}

For a given student $s$ and course $c$ pair, we can discover multiple paths from student $s$ to other student $s'$ by their common tags in the $\mathcal{KG}$. By concatenating these paths with the paths from student $s'$ to course $c$, we define such student-tag-student-course paths (e.g., the solid links in figure \ref{student_profiles}) as $Similar\,Student\,Paths\,(SSP)$. We leverage such paths between two students to measure whether they are similar. The multiple-step paths with the pattern like student-course-course, which contain the prior course knowledge (e.g., the dashed links in figure \ref{student_profiles}), are defined as $Course\,Knowledge\,Paths\,(CKP)$.

Formally, the $SSP$ between a given student $s$ and course $c$ can be represented as a path set $\mathcal{P}_s=\{P^s_1,P^s_2,\ldots,P^s_M\}$ where $P^s_M = \{p_1,p_2,\ldots,p_K\}$ is a path set between two similar students. And $M,K$ denotes the number of similar students and paths between two students. Similarly, the paths between courses defined as a path set $\mathcal{P}_c=\{P^c_1,P^c_2,\ldots,P^c_N\}$ where $N$ denotes the number of courses which related to course $c$. Thus, we define the path set between student and course as $\mathcal{P}_{s,c}=\mathcal{P}_s\cup\mathcal{P}_c$. The sequences in each path can be detailed as $p=[(v_1,n_1,r_1) \xrightarrow{} (v_2,n_2,r_2) \xrightarrow{} \ldots \xrightarrow{} (v_L,n_L,<End>)]$, where $v_l$ and $n_l$ separately denote the value and node type of the entity in path $p$, $r_l$ is the relationship between $n_{l}$ and $n_{l+1}$.

In this end, we give a real example to explain how humans use $SSP$ and $CKP$ to predict students' performance. We formalize the paths in figure \ref{student_profiles}, where student Bob and student Tom have several same tags in different semesters.

\begin{itemize}
    \item $p_1=[(Bob,Student,\textbf{have}) \xrightarrow{} (1,Grind,\textbf{belong\_to})\\ \xrightarrow{} (Tom,Student,\textbf{get\_score}) \xrightarrow{} (Pass,Score,\textbf{in}) \xrightarrow{}(OS,Course,<End>)] \xrightarrow{\text{Semantic\,Transformation}}$\\
    Bob had a "Grind" tag which also belonged to Tom, who passed theOS, means that Bob and Tom are both excellent

    \item $p_2=[(Bob,Student,\textbf{get\_score}) \xrightarrow{} (Pass,Score,\textbf{in}\\) \xrightarrow{} (DS, Course,\textbf{prerequisite}) \xrightarrow{} (OS, Course,<End>)] \xrightarrow{\text{Semantic\,Transformation}}$\\
    Bob passed the DS, which is the OS's prerequisite course, so maybe Bob can still perform well in the OS.
\end{itemize}

Based on the principle of collaborative filtering that similar students will get similar achievements. We have reasons to infer that Bob will pass the OS because most of the similar students have passed and is good at the prerequisite course of the OS.

\subsection{Modeling process of ESPA}
\begin{figure}[h]
    \centering
    \includegraphics[width=0.8\linewidth]{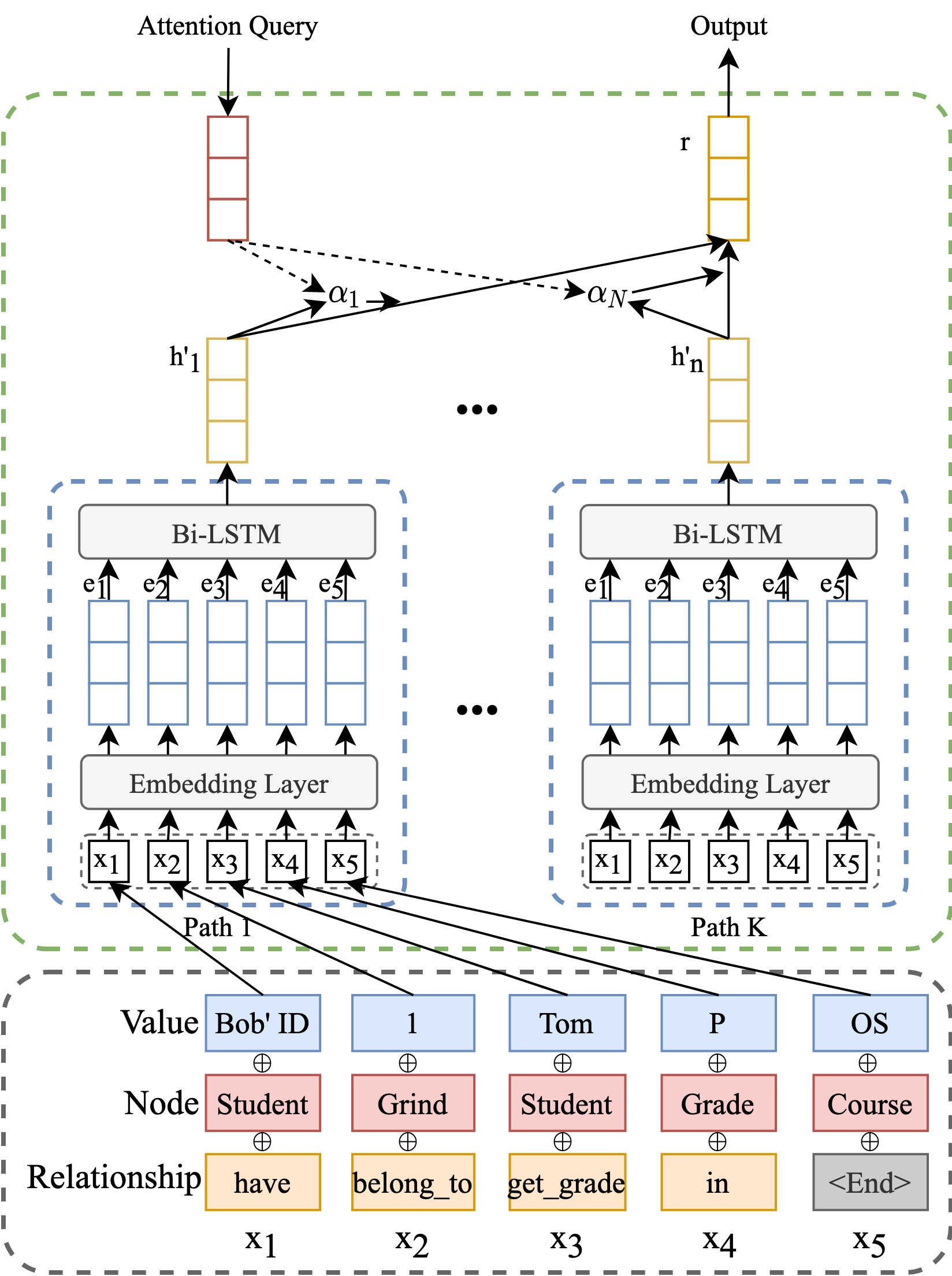}
    \caption{The student path encoder with personalized attention.}
    \label{path_encoder}
\end{figure}

\paragraph{Path Encoder} In order to measure the similarity between two students, we propose the path encoder shown in figure \ref{path_encoder} to integrate information for all paths in $SSP$. We leverage typical Long Short-term Memory (LSTM) network to learn the semantic representations of each path $p_i$ from input shown as figure \ref{path_encoder}. First, we contact the triplet's embeddings as input to each step of LSTM $e_t$ for the path-step $t$.
\begin{equation}
    e_t=e_v \oplus e_n \oplus e_r ,
\end{equation}
where $e_t\in\mathbb{R}^{3D_e}$, $e_v$, $e_n$, $e_r\in\mathbb{R}^{D_e}$ are the embeddings of entity value, entity type, relationships, and $D_e$ denotes the dimension of the embedding.
In this way, the input of each time step contains the information of the nodes and relationships. Consequently, $h_{t-1}$ and $e_t$ are used to learn the hidden state of each path-step in path $p_i$, which is defined as the following equations:
\begin{equation}
    \begin{aligned}
      f_t&=\sigma(W_fe_t+U_f{h_{t-1}}+b_f),\\
      i_t&=\sigma(W_ie_t+U_i{h_{t-1}}+b_i),\\
      o_t&=\sigma(W_oe_t+U_o{h_{t-1}}+b_o),\\
      \tilde{c_t}&=tanh(W_ce_t+U_c{h_{t-1}}+b_c),\\
      c_t&=f_t \odot c_{t-1} +i_t \odot \tilde{c_t},\\
      h_t&=o_t \odot \sigma(c_t),
    \end{aligned}
\end{equation}
where $c_t\in\mathbb{R}^{D_h}$ and $\tilde{c_t}\in\mathbb{R}^{D_h}$ denote the cell state and information transform module, and $D_h$ is the number of hidden units; $f_t$, $i_t$, and $o_t$ separately represent the forget, input, and output gate. $W_*\in\mathbb{R}^{D_h\times 3D_e}$,$U_*\in\mathbb{R}^{D_h\times D_h}$ and $b_*\in\mathbb{R}^{D_h}$ are the weight parameters in LSTM. Particularly, $\sigma(\cdot)$ is the sigmoid activation function and $\odot$ stands for the element-wise product of two vectors.

The traditional LSTM model learns each step represented by a single direction network and connot utilize the contextual information from the future step during the training process\cite{TanXZ15}. Since both head and tail information of the path are significant, we build a BiLSTM-based network, taking the path sequence in both forward and backward directions. Thus we get the full path semantic information $h'_n\in\mathbb{R}^{2D_h}$ with the concatenation of bidirectional hidden state in the last step.

Aforementioned, each student's academic performance is various, even in the same living habits. Such we introduce a novel personalized attention mechanism to capture this difference. Because the $SSP$ contains the embedding of the student ID, we denote the student ID embedding as the query of the dot-product attention mechanism\cite{VaswaniSPUJGKP17} for more efficient parameter updates. We use a dense layer to learn the local-level student preference query $q_l$ as:
\begin{equation}
    q_l=ReLU(W_le_s+b_l),
\end{equation}
where $W_l\in \mathbb{R}^{2D_h\times D_e}$ and $b_l \in \mathbb{R}^{2D_h}$ are parameters, $2D_h$ is the query size. In this module, each path's attention weight is calculated based on the interactions between the local-level query and path representations. We denote the attention weight of the path i as $\alpha_i$, which formulated as:
\begin{equation}
    \begin{aligned}
      a'_i&=\frac{q_l {h'_i}^T}{\sqrt{d_k}},\\
      \alpha'_i&=\frac{exp(a'_i)}{\sum_{j=1}^K{exp(a'_j)}},\\
    \end{aligned}
    \label{equ:local_attention_score}
\end{equation}
where $d_k=D_e$ is used to prevent the dot product from being too large.

The output of the path encoder $r_i$ is the summation of the contextual representations of paths weighted by their attention score:
\begin{equation}
    r_i=\sum_{j=1}^K{\alpha'_jh'_j},
  \end{equation}

As for $CKP$, we use the same encoder to capture the representations of different paths between courses.

\paragraph{Predict with Personalized Attention} Cause the similarity between students is different, and the impact of related courses on the course to be predicted is also different. Thus, we apply the personalized attention mechanism to each path encoder's output. The global-level attention query $q_g$ for the output from each path encoder still learned from a dense layer:
\begin{equation}
    q_g=ReLU(W_ge_s+b_g),
\end{equation}
where $W_l\in \mathbb{R}^{2D_h\times D_e}$ and $b_l \in \mathbb{R}^{2D_h}$ are parameters, $2D_h$ is the dimention of global-level attention query. This query $q_g$ represents sutdent's long-term learning status.

For global-level attention, the attention weight of each representation $r_i$ is formulated as:
\begin{equation}
    \begin{aligned}
      a_i&=\frac{\beta_i q_gr_i^T}{\sqrt{d_k}},\\
      \alpha_i&=\frac{exp(a_i)}{\sum_{j=1}^{M+N}{exp(a_j)}},\\
    \end{aligned}
    \label{equ:global_attention_score}
\end{equation}
where $\beta_i$ denotes that there are different weights for $SSP$ and $CKP$.

As for the prediction, we use the embedding of grade tag $e'_i$ (right part in the figure \ref{model}), corresponding to each $r_i$, as a value for the global-level attention mechanism. Note that $e'_i$ may equal to $e_v$ cause the grade tag is also included in $SSP$ and $CKP$. Therefore the model learns better grade representation.
\begin{equation}
    v_i=tanh(W_ve'_i+b_v),
\end{equation}
where $W_v\in \mathbb{R}^{1 \times D_e}$ is the projection weight for $v_i\in \mathbb{R}$.

We notice that the average score for each course was different due to human factors (such as teachers' scoring habits). At the same time, each student has a different average score because of different learning foundations. Following the prior work of \cite{Koren08}, we leverage two biases to represent these two cases separately. The probability $\hat{y}'_{s,c}$ of the student $s$ may fail in the course $c$ is calculated by the inner product with two biases first, and then activated by the sigmoid function, which is formulated as:
\begin{equation}
    \hat{y}_{s,c}=\sigma(\sum_i^{M+N}\alpha_iv_i+b_s+b_c),
\end{equation}
where $b_s$ and $b_c$ are learning parameters for each student and course. And $M+N$ is the number of paths set $\mathcal{P}_{s,c}$ for a given student $s$ and course $c$.

\subsection{Model learning} The loss function of ESPA is the negative log-likelihood of the observed sequences between student to course. Formally, for predicting whether student $s$ fail on course $c$, $\hat{y}_{s,c}$ is the predicted result from the model and ${y_{s,c}}$ is the ground truth. Thus the loss for student performance prediction is defined as:
\begin{equation}
    loss_{pred}=-[y_{s,c}\log(\hat{y}_{s,c})+(1-y_{s,c})\log(1-\hat{y}_{s,c})]
\end{equation}

\subsection{Additional Inference Task} To ensure that the cosine distances of similar students' embeddings can be updated to closer, we design a subtask for better parameter learning inspired by prior works \cite{LianYZLXX16,ZhangSPST17,YaoLCWZ19}. We utilize a shallow neural network to predict each similar student $Student 1..M$ in $\mathcal{P}_s$ for a given student $s$. Thus, we get better student representations to assist the main task. The prediction process is as follows:
\begin{equation}
    P(s_j|s)=softmax(W_se_s+b_s)
\end{equation}
where $W_s\in \mathbb{R}^{S' \times D_e}$ and $b_s \in \mathbb{R}^{S'}$ are the weight parameters and bias of the layer respectively. And $P(s_j|s)$ denotes the posterior probability that $s_j$ is a similar student of student $s$. To this end, the loss function of the subtask and the integrated loss function of our model is defined as:
\begin{equation}
    loss_{infe}=-\sum^M_{i=1}\sum^{S'}_{j=1}y'_{s,c,j}log(P(s_j|s))
\end{equation}
\begin{equation}
    L=\frac{1}{N}\sum_{n=1}^N(loss_{pred}+\lambda loss_{infe})
\end{equation}
where $S'$ and $M$ are the total numbers of students and the number of similar students of student $s$, separately. $N$ is the number of samples. $y'_{s,c,j}$ denotes whether $s_j$ is a similar student of $s$. And $\lambda$ controls the trade-off between the performance prediction loss and the subtask loss.
\section{Experiments}
In this section, we conduct extensive experiments to demonstrate ESPA's effectiveness from these two aspects: (1) the prediction performance of ESPA against the baselines; (2)the explainability of the model throughout the case study.

\subsection{Experimental Dataset}

We applied for students' learning and behavior data from the Information Department of a college due to the lack of a multi-source public dataset to model the long short-term student profile. Observed school card information starts in Spring 2015 and continues until Fall 2018. We filtered students from grades 2013 to 2016 from three majors, where there are 2,409 students. During this period, these students had taken 590 unique courses, 126,454 score records, and 4,628,802 card records (e.g., canteen consumption, bathing, shopping). 

Students were modeled by analyzing student card data with data mining methods. We analyzed the student's learning status and behavioral habits in each semester and the whole semesters. The tags in the student profile are shown in table \ref{tab:tags}. We also crawled the course information from a MOOC website to build the coursess knowledge graph. In the end, we integrated the student profile, necessary information (e.g., statistical information, academic information), and course knowledge into the $\mathcal{KG}$.

\begin{table}
    \begin{adjustwidth}{0cm}{0cm}
      \begin{tabular}{p{4.5cm}p{2cm}p{1cm}}
        \toprule[1pt]
        description & tags & period\\
        \hline
        Based on the upper and lower quartiles of a student's overall academic performance.  & Grind, Ordinary, Slacker & long term\\
        \hline
        An overall student failing based on the number of the courses they failed.   & None, Few, Repeat Risk, Drop out Risk & long term\\
        \hline
        Changes in student rankings in different semesters reflect students' learning status trends. & Ascend, Descend & short term\\
        \hline
        Different semesters' dining habits data reveals a healthy diet benefits academic development. & Dietary, Regular, Irregular & short term\\
        \hline
        Based on the number of breakfasts each month, breakfast affects student learning status typically. & Breakfast Habit, No Breakfast Habit & short term\\
        \hline
        Sleep time approximately calculated from campus gateway data. & Sleep Late, Sleep on Time & short term\\
        \hline
        Consumption situation based on the upper and lower quartiles of students' consumption.  & Low, Normal, High Consumption & short term\\
        \bottomrule[1pt]
      \end{tabular}
      \end{adjustwidth}
      \caption{The tags in the student profile}
      \label{tab:tags}
\end{table}

\subsection{Path Selection}
The number of nodes in the $\mathcal{KG}$ is 9,755, which generated 569,738 relationships. Thus, it is infeasible to fully exploring all connected paths over the $\mathcal{KG}$. As pointed out by \cite{SunHYYW11}, paths with length greater than six will introduce noisy entities. After analyzed the average length of the student-course pairs, we used a specific pattern to sample the paths in the $\mathcal{KG}$, each with a length up to five. For a student $s$, we filtered out 60 similar students with more than or equal to one path between them. In the end, the average number of paths between two similar students is five.

\subsection{Experimental Settings}
\paragraph{Evaluation Metrics} We adopted these evaluation protocols to evaluate the performance of predicting student performance, give by:
\begin{itemize}
  \item \textbf{precision} refers to the closeness of the measurements to a specific value. It is used to measure the correct number of predicted samples.
  \item \textbf{recall} also known as sensitivity, which is the fraction of the total amount of relevant retrieved instances.
  \item \textbf{f1-score} considers both the precision and the recall, which is the harmonic mean of precision and recall.
  \item \textbf{AUC} tells how much the model is capable of distinguishing between classes. The larger the value of AUC, the better the effect of the model.
\end{itemize}

\paragraph{Baselines} 
We compared ESPA with SVD++ \cite{Koren08}, NCF \cite{HeLZNHC17}, KPRN \cite{WangWX00C19}, XGboost, DeepFM \cite{GuoTYLH17}, xDeepFM \cite{LianZZCXS18}, AutoInt \cite{SongS0DX0T19}. As introduced and discussed prior, these models are related to our task, and some are state-of-the-art methods. 
The aforementioned works (e.g., CritNet \cite{KimVG18, KimVG182}, EKT \cite{LiuHYCXSH19}) only considered the students' exercises. However, we are only concerned about the influence of student behavior and relationships between courses on performance. Due to different data formats, the related comparisons will not be conducted.
\paragraph{Parameter Settings} During the training process, the orthogonal matrics were used to initialize the LSTM and Xavier normalization to initialize all linear layer's weight parameters. We optimized all parameters with Adam \cite{KingmaB14} and used the grid search strategy to find out the best sets of hyperparameters. The learning rate was searched in \{0.02, 0.01, 0.001, 0.0001\}, and batch size in \{128, 256, 512 ,1024\}. Other hyperparameters are as follows: the embedding size of students, courses, tags, and its values was 16,  considering the total number of entities. Moreover, the hidden size of BiLSTM was 12. We founded that setting the type weight $\beta$ of $SSP$ higher yielded better results. $\beta_s:\beta_c=0.7:0.3$ is the best, which shows that behavior has a greater impact on student performance. Furthermore, we set the trade-off parameter $\lambda$ as 1 in our experiments.

\begin{table}
    \begin{adjustwidth}{0cm}{0cm}
    \begin{tabular}{p{1.4cm}p{1.2cm}p{1cm}p{0.8cm}p{0.7cm}p{0.7cm}}
      \toprule[1pt]
      Method& Target & Precision & Recall & F1-score & AUC\\
      \hline
      \multirow{2}*{SVD++} & 0 (pass) & 0.94& 0.98 &0.96 & \multirow{2}*{0.76}\\
      ~ & 1 (fail) & \textbf{0.52} & 0.19 & 0.28 & ~\\
      \hline
      \multirow{2}*{NCF} & 0 (pass) & 0.96 & 0.80 & 0.87 &\multirow{2}*{0.81}\\
      ~ & 1 (fail) & 0.23 & 0.66 & 0.34 & ~\\
      \hline
      \multirow{2}*{KPRN} & 0 (pass) & 0.97 & 0.84 & 0.90 &\multirow{2}*{0.83}\\
      ~ & 1 (fail) & 0.26 & 0.67 & 0.39 &~\\
      \hline
      \multirow{2}*{KPRN+} & 0 (pass) & 0.98 & 0.86&0.92 &\multirow{2}*{0.85}\\
      ~ & 1 (fail) & 0.30 & 0.72 & 0.42 &~\\
      \hline
      \multirow{2}*{XGboost} & 0 (pass) &  0.98 & 0.84 & 0.90 & \multirow{2}*{0.83}\\
      ~ & 1 (fail) & 0.37 & 0.83 & 0.51 & ~\\
      \hline
      \multirow{2}*{DeepFM} & 0 (pass) & 0.97  & 0.79 & 0.87 & \multirow{2}*{0.77}\\
      ~ & 1 (fail) & 0.23 & 0.75 & 0.36 & ~ \\
      \hline
      \multirow{2}*{xDeepFM} & 0 (pass) & 0.96  & 0.81 & 0.88 & \multirow{2}*{0.74}\\
      ~ & 1 (fail) & 0.23 & 0.66 & 0.35 & ~ \\
      \hline
      \multirow{2}*{AutoInt} & 0 (pass) & 0.97  & 0.84 & 0.90 & \multirow{2}*{0.79}\\
      ~ & 1 (fail) & 0.25 & 0.70 & 0.37 & ~ \\
      \hline
      \textbf{\multirow{2}*{ESPA}} & 0 (pass) &  0.98 & 0.90 & 0.94 & \multirow{2}*{\textbf{0.93}} \\
      ~ & 1 (fail) & 0.42 & \textbf{0.87} & \textbf{0.57}&~\\
      \hline
      \textbf{\multirow{2}*{w/o biases}} & 0 (pass) &  0.98 & 0.90 & 0.93 & \multirow{2}*{0.91} \\
      ~ & 1 (fail) & 0.39 & 0.75 & 0.51 &~\\
      \hline
      \textbf{\multirow{2}*{w/o subtask}} & 0 (pass) &  0.97 & 0.89 & 0.93 & \multirow{2}*{0.90} \\
      ~ & 1 (fail) & 0.37 & 0.74 & 0.49 &~\\
      \bottomrule[1pt]
  \end{tabular}
  \caption{Comparison results on the test set using the precision, recall and f1-score.}
    \label{compare}
  \end{adjustwidth}
\end{table}

\subsection{Performance Comparison}
\paragraph{Student Performance Prediction}
To simulate the real situation, we filtered all data in Fall 2018, which belongs to grade 2016 students in a major, to construct the testing set and rest data for constructing the training set. Such a division method can prevent the problem of information leakage during the training process. We fed the original statistical data used to construct the $KG$ to other competitors. For a fair comparison, we trained our model without any pre-trained embeddings. We also did grid searches for the baseline algorithms to ensure that each baseline achieves state-of-the-art performance.

Furthermore, we replaced the pooling layer with a dot-product attention network in KPRN and denoted it as KPRN+, which is not implied in the paper. It is worth noting that the label shows a significant imbalance, where the number of fail records is much less. We had balanced data for all the methods, such as downsampling, weighted loss function.

Table \ref{compare} reports our experimental results. It is worth focusing on the effects of each method to predict failure grades. Moreover, our model achieved state-of-the-art performance as for recall, f1-score, and AUC in all methods. According to the results, We have several observations.

First, the deep methods (e.g., NCF, KPRN, and ESPA) outperform traditional matrix factorization. That is because neural networks can learn more sophisticated features than SVD++, which helps learn more informative latent factors of entities and relationships. 

Second, it shows that the deep learning methods using attention mechanism (KPRN+, ESPA) outperform most of the methods without attention mechanism (KPRN, DeepFM, xDeepFM). This phenomenon is because different student preferences and courses have different informativeness for student performance. It is difficult for a neural network without an attention mechanism to capture this personalized difference. Furthermore, the result also shows that it is worthwhile to model the student profile and course knowledge graph explicitly.

Third, compared to KPRN, our approach is more suitable for users with many connections between them (e.g., college student profile). KPRN uses each path between items to learn users' preferences and other representations. However, we focus on the similarity between the two students or courses in higher education. By studying multiple paths between students and combining the local- and global- level attention mechanisms, we got state-of-the-art results.

Finally, we evaluated the effectiveness of student and course biases and the subtask shown in table \ref{compare}. We found that these two biases can improve model performance but are not decisive. Therefore, they do not significantly affect the explainability of the results. Also, we found that the subtask can help the model learn the student representations in the direction we expect.

\begin{figure}
    \centering
    \includegraphics[width=0.9\linewidth]{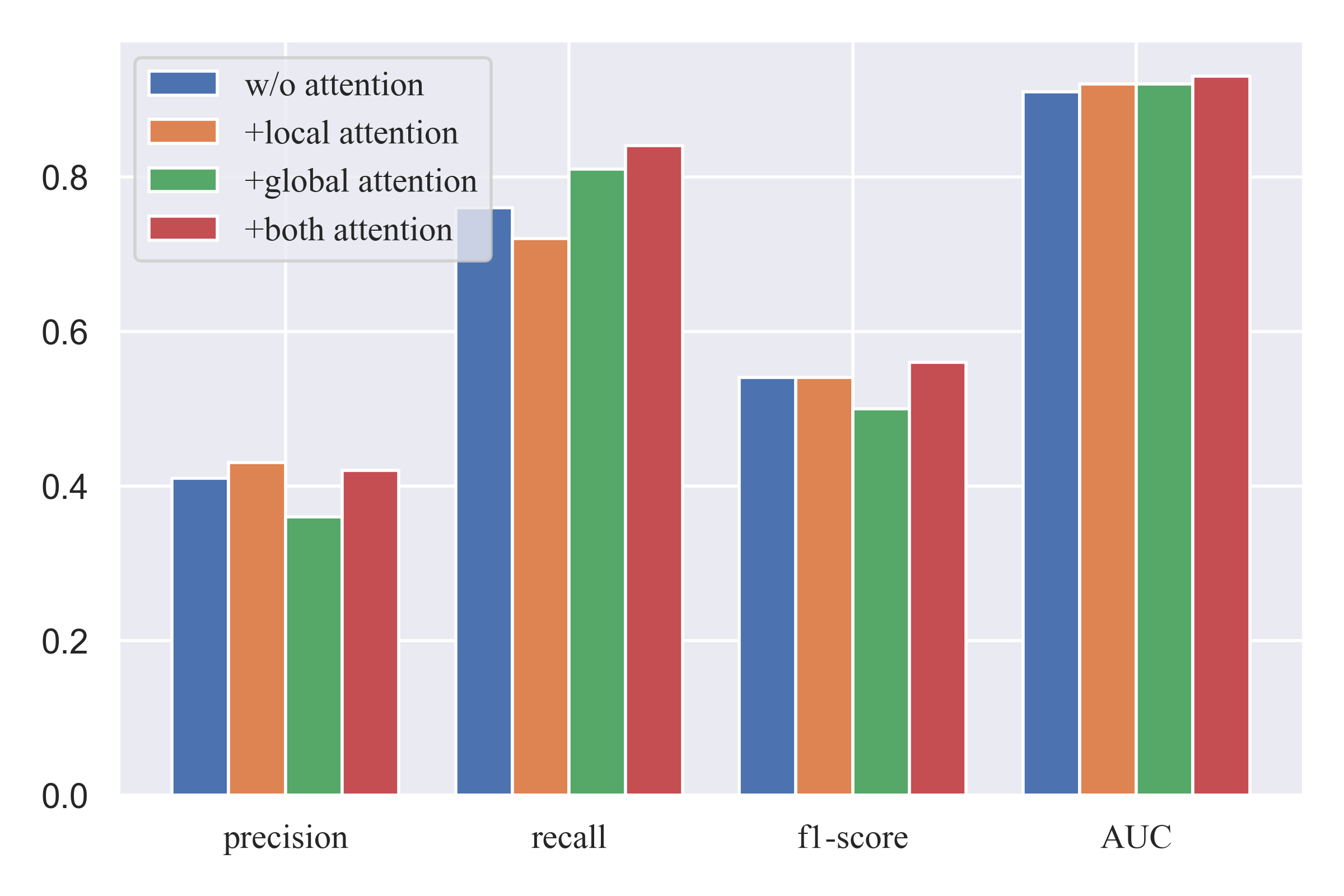}
    \caption{ The effectiveness of the personalized attention network. }
    \label{compare_attention}
\end{figure}

\paragraph{Ablation Experiment} In this section, we conducted several experiments to explore the personalized attention mechanism's effectiveness in our approach. We replaced personalized attention with weighted sum pooling. According to figure \ref{compare_attention}, we have several observations.

First, local-level personalized attention can effectively improve the performance of our approach. It is because paths are basic units to convey information about students' behavior and performance. Moreover, selecting the remarkable paths according to student preferences is useful for learning more informative paths representations when predicting performance.

Second, global-level personalized attention can also improve the performance of our model. Cause representations of similar students and courses usually have different informativeness for learning student representations. Recognizing the critical students and course is significant for learning high-quality student representations and predicting.

\subsection{Case Study} 
\begin{figure}[h]
    \includegraphics[width=0.9\linewidth]{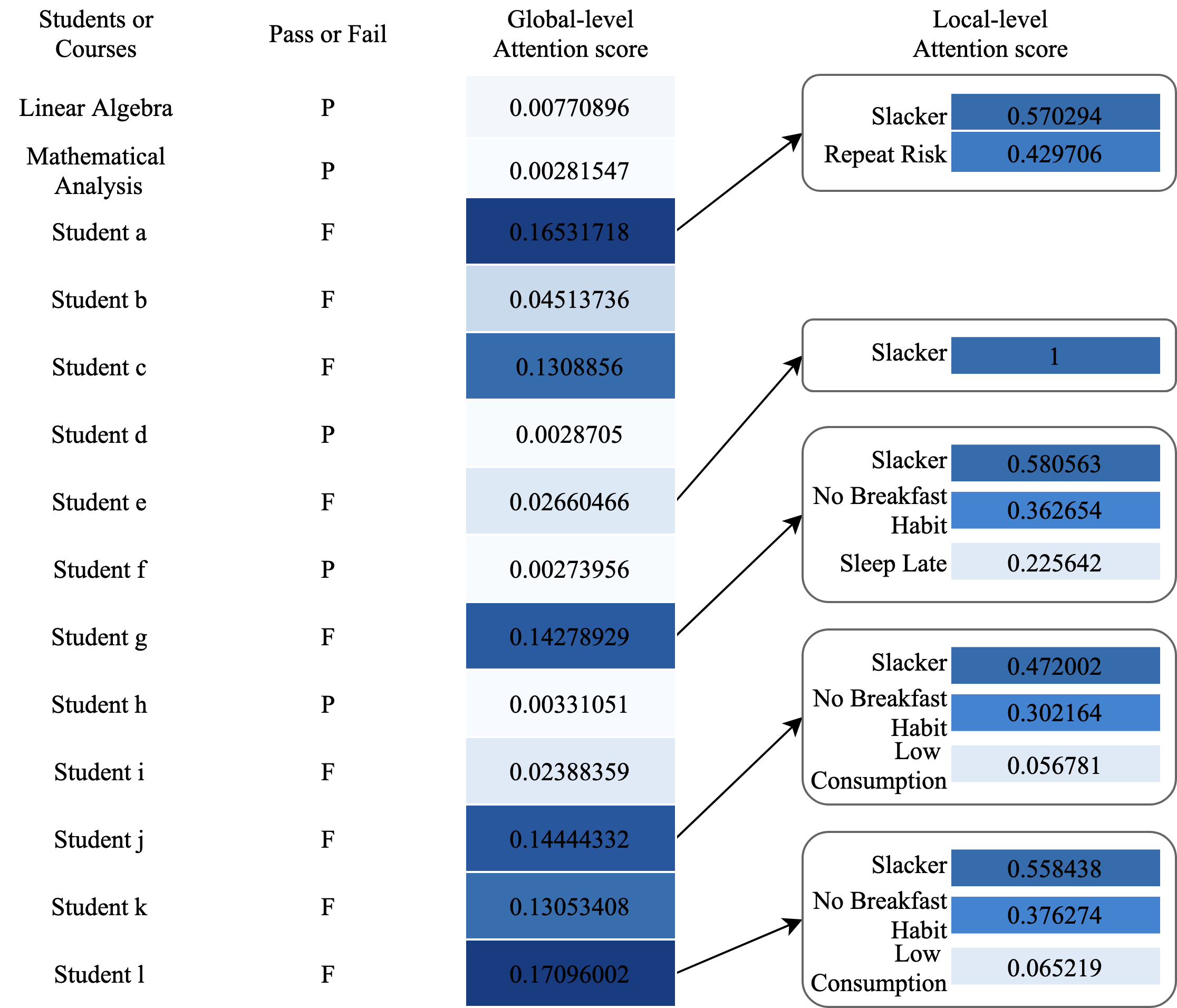}
    \caption{Illustrate the attention distribution for paths within a given student-course pair.}
    \label{attention}
\end{figure}

To improve the confidence of our model, we explored the explainability of the predicted results. We randomly selected a student-course pair $\mathcal{P}_{s,c}$ for evaluating. The local-level and the global-level attention score of the student ${s}$ are shown in figure \ref{attention}. 

\begin{figure}[h]
    \includegraphics[width=\linewidth]{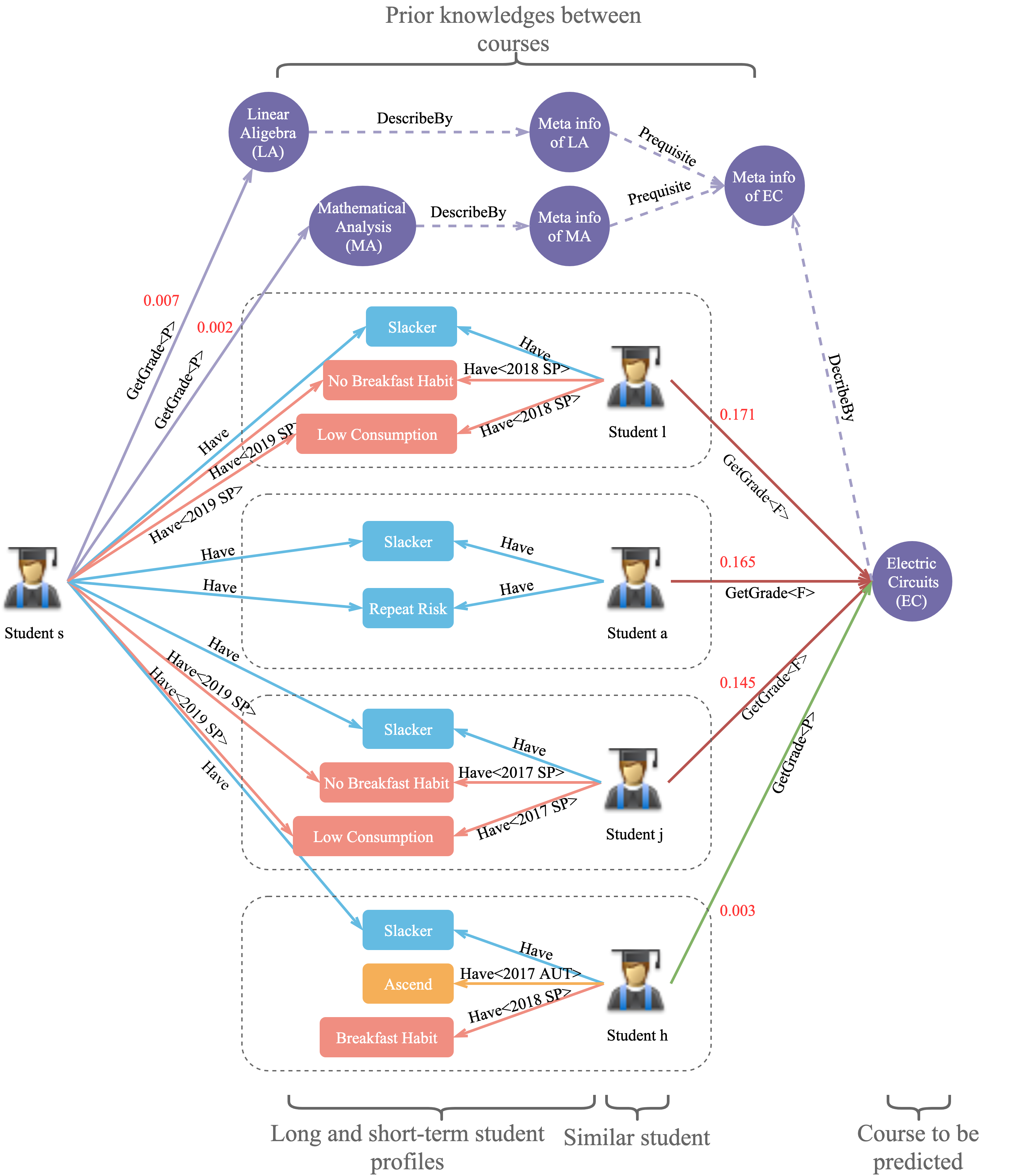}
    \caption{A case study of highest attention paths in knowledge graph.}
    \label{case_study}
\end{figure}

Our model correctly predicted that the student $s$ would fail in Electric Circuits (EC). It can be seen intuitively from figure \ref{attention} that most of the students who are similar to student $s$ are failed in the EC. Throughout the global-level attention scores, we found that most students with high attention scores failed the exam. Which is the main reason that model predicted student $s$ might fail in the EC. It is worth noting why the attention scores of student $b$,$e$,$i$, who either failed in the course, are lower. Because of the personalized attention mechanism, the model learned that student $b$,$e$,$i$ were not similar to the student $s$. For example, there is only one path between student $e$ and student $s$. 

Then we explored the relationships between student $s$ and students with high attention scores. It can be concluded from figure \ref{case_study} that the student $s$'s life was irregular throughout the course, while students with similar habits either failed in the course. Hence, the model predicted that the student's failure was reasonable. From the student profile, we can intuitively understand why student $h$'s attention score is much lower than others. Although student $h$ also had a Slacker tag, the student $h$ had some more active tags such as Ascend, Breakfast habit. These tags' information was not modeled in the paths between student $s$ and student $h$ explicitly, but such information was updated to the embedding of student $h$ during the entire model training process. We believe the model could understand the representations of students, tags, and courses in the paths for decisions.

When we apply the model to the real scenario, educators are more concerned about which behaviors affected student performance rather than attention scores. Thus we will highlight essential tags based on attention scores, such as No Breakfast habit, Low Consumption. At the same time, we will show them student profiles of similar students, such as figure \ref{case_study}. Educators can utilize the above information to intervene in students in advance and improve their living habits and grades. Simultaneously, educators can intuitively see the basis of model inferring, thereby increasing confidence in model results.

\section{Conclusions}
In this paper, we mapped the student performance prediction problem in education to the recommendation system. The ESPA model captures the semantic information of the paths between students and courses in the knowledge graph. Besides, our model achieves state-of-the-art results and has explainability under our designed attention mechanism. We did not emphasize the student profile's construction method because various tags can be added to student profiles in practical applications. Simultaneously, based on this method, we can also add the relationships between teachers and courses.

In the future, we will extend our work in these directions. First, we found that the real environment data is exceptionally imbalanced, where the number of students passing the course is far more than that of the students who fail. However, most of the studies have ignored this problem. Thus, we expect to use methods similar to anomaly detection for predicting failure results. We would also like to build up more accurate and timely modeling of students for efficient path representation. Meanwhile, we willing attempt to perform a holistic analysis of the student-student-course using an improved RNN structure or a graph neural network.

\section{Acknowledgments}
This work is supported by the National Key Research and Development Program of China (2016YFE0204500), The work was supported by the National Natural Science Founda- tion of China (Grant No.61971066) and the Beijing Natural Science Foundation (No. L182038), the Fundamental Research Funds for the Central Universities and Beijing University of Posts and Telecommunications 2017 Education and Teaching Reform Project No.2017JY31.

\bibliographystyle{aaai}
\bibliography{references}

\begin{thebibliography}{}

\bibitem[\protect\citeauthoryear{Abdelrahman and Wang}{2019}]{AbdelrahmanW19}
Abdelrahman, G., and Wang, Q.
\newblock 2019.
\newblock Knowledge tracing with sequential key-value memory networks.
\newblock In Piwowarski, B.; Chevalier, M.; Gaussier, {\'{E}}.; Maarek, Y.;
  Nie, J.; and Scholer, F., eds., {\em Proceedings of the 42nd International
  {ACM} {SIGIR} Conference on Research and Development in Information
  Retrieval, {SIGIR} 2019, Paris, France, July 21-25, 2019},  175--184.
\newblock {ACM}.

\bibitem[\protect\citeauthoryear{Ai \bgroup et al\mbox.\egroup
  }{2019}]{AiCGZWFW19}
Ai, F.; Chen, Y.; Guo, Y.; Zhao, Y.; Wang, Z.; Fu, G.; and Wang, G.
\newblock 2019.
\newblock Concept-aware deep knowledge tracing and exercise recommendation in
  an online learning system.
\newblock In Desmarais, M.~C.; Lynch, C.~F.; Merceron, A.; and Nkambou, R.,
  eds., {\em Proceedings of the 12th International Conference on Educational
  Data Mining, {EDM} 2019, Montr{\'{e}}al, Canada, July 2-5, 2019}.
\newblock International Educational Data Mining Society {(IEDMS)}.

\bibitem[\protect\citeauthoryear{Bydzovsk{\'{a}}}{2015}]{Bydzovska15}
Bydzovsk{\'{a}}, H.
\newblock 2015.
\newblock Are collaborative filtering methods suitable for student performance
  prediction?
\newblock In Pereira, F.~C.; Machado, P.; Costa, E.; and Cardoso, A., eds.,
  {\em Progress in Artificial Intelligence - 17th Portuguese Conference on
  Artificial Intelligence, {EPIA} 2015, Coimbra, Portugal, September 8-11,
  2015. Proceedings}, volume 9273 of {\em Lecture Notes in Computer Science},
  425--430.
\newblock Springer.

\bibitem[\protect\citeauthoryear{Catherine \bgroup et al\mbox.\egroup
  }{2017}]{CatherineMEC17}
Catherine, R.; Mazaitis, K.; Esk{\'{e}}nazi, M.; and Cohen, W.~W.
\newblock 2017.
\newblock Explainable entity-based recommendations with knowledge graphs.
\newblock In Tikk, D., and Pu, P., eds., {\em Proceedings of the Poster Track
  of the 11th {ACM} Conference on Recommender Systems (RecSys 2017), Como,
  Italy, August 28, 2017}, volume 1905 of {\em {CEUR} Workshop Proceedings}.
\newblock CEUR-WS.org.

\bibitem[\protect\citeauthoryear{Chaudhari, Azaria, and
  Mitchell}{2017}]{ChaudhariAM17}
Chaudhari, S.; Azaria, A.; and Mitchell, T.~M.
\newblock 2017.
\newblock An entity graph based recommender system.
\newblock {\em {AI} Commun.} 30(2):141--149.

\bibitem[\protect\citeauthoryear{Dascalu \bgroup et al\mbox.\egroup
  }{2016}]{DascaluPBCT16}
Dascalu, M.; Popescu, E.; Becheru, A.; Crossley, S.~A.; and Trausan{-}Matu, S.
\newblock 2016.
\newblock Predicting academic performance based on students' blog and microblog
  posts.
\newblock In Verbert, K.; Sharples, M.; and Klobucar, T., eds., {\em Adaptive
  and Adaptable Learning - 11th European Conference on Technology Enhanced
  Learning, {EC-TEL} 2016, Lyon, France, September 13-16, 2016, Proceedings},
  volume 9891 of {\em Lecture Notes in Computer Science},  370--376.
\newblock Springer.

\bibitem[\protect\citeauthoryear{Dutt, Ismail, and Herawan}{2017}]{DuttIH17}
Dutt, A.; Ismail, M.~A.; and Herawan, T.
\newblock 2017.
\newblock A systematic review on educational data mining.
\newblock {\em {IEEE} Access} 5:15991--16005.

\bibitem[\protect\citeauthoryear{Guo \bgroup et al\mbox.\egroup
  }{2017}]{GuoTYLH17}
Guo, H.; Tang, R.; Ye, Y.; Li, Z.; and He, X.
\newblock 2017.
\newblock Deepfm: {A} factorization-machine based neural network for {CTR}
  prediction.
\newblock In Sierra, C., ed., {\em Proceedings of the Twenty-Sixth
  International Joint Conference on Artificial Intelligence, {IJCAI} 2017,
  Melbourne, Australia, August 19-25, 2017},  1725--1731.
\newblock ijcai.org.

\bibitem[\protect\citeauthoryear{He \bgroup et al\mbox.\egroup
  }{2017}]{HeLZNHC17}
He, X.; Liao, L.; Zhang, H.; Nie, L.; Hu, X.; and Chua, T.
\newblock 2017.
\newblock Neural collaborative filtering.
\newblock In Barrett, R.; Cummings, R.; Agichtein, E.; and Gabrilovich, E.,
  eds., {\em Proceedings of the 26th International Conference on World Wide
  Web, {WWW} 2017, Perth, Australia, April 3-7, 2017},  173--182.
\newblock {ACM}.

\bibitem[\protect\citeauthoryear{Huang \bgroup et al\mbox.\egroup
  }{2018}]{HuangZDWC18}
Huang, J.; Zhao, W.~X.; Dou, H.; Wen, J.; and Chang, E.~Y.
\newblock 2018.
\newblock Improving sequential recommendation with knowledge-enhanced memory
  networks.
\newblock In Collins{-}Thompson, K.; Mei, Q.; Davison, B.~D.; Liu, Y.; and
  Yilmaz, E., eds., {\em The 41st International {ACM} {SIGIR} Conference on
  Research {\&} Development in Information Retrieval, {SIGIR} 2018, Ann Arbor,
  MI, USA, July 08-12, 2018},  505--514.
\newblock {ACM}.

\bibitem[\protect\citeauthoryear{Kim, Vizitei, and Ganapathi}{2018a}]{KimVG182}
Kim, B.; Vizitei, E.; and Ganapathi, V.
\newblock 2018a.
\newblock Gritnet 2: Real-time student performance prediction with domain
  adaptation.
\newblock {\em CoRR} abs/1809.06686.

\bibitem[\protect\citeauthoryear{Kim, Vizitei, and Ganapathi}{2018b}]{KimVG18}
Kim, B.; Vizitei, E.; and Ganapathi, V.
\newblock 2018b.
\newblock Gritnet: Student performance prediction with deep learning.
\newblock In Boyer, K.~E., and Yudelson, M., eds., {\em Proceedings of the 11th
  International Conference on Educational Data Mining, {EDM} 2018, Buffalo, NY,
  USA, July 15-18, 2018}.
\newblock International Educational Data Mining Society {(IEDMS)}.

\bibitem[\protect\citeauthoryear{Kingma and Ba}{2015}]{KingmaB14}
Kingma, D.~P., and Ba, J.
\newblock 2015.
\newblock Adam: {A} method for stochastic optimization.
\newblock In Bengio, Y., and LeCun, Y., eds., {\em 3rd International Conference
  on Learning Representations, {ICLR} 2015, San Diego, CA, USA, May 7-9, 2015,
  Conference Track Proceedings}.

\bibitem[\protect\citeauthoryear{Koren}{2008}]{Koren08}
Koren, Y.
\newblock 2008.
\newblock Factorization meets the neighborhood: a multifaceted collaborative
  filtering model.
\newblock In Li, Y.; Liu, B.; and Sarawagi, S., eds., {\em Proceedings of the
  14th {ACM} {SIGKDD} International Conference on Knowledge Discovery and Data
  Mining, Las Vegas, Nevada, USA, August 24-27, 2008},  426--434.
\newblock {ACM}.

\bibitem[\protect\citeauthoryear{Lian \bgroup et al\mbox.\egroup
  }{2016}]{LianYZLXX16}
Lian, D.; Ye, Y.; Zhu, W.; Liu, Q.; Xie, X.; and Xiong, H.
\newblock 2016.
\newblock Mutual reinforcement of academic performance prediction and library
  book recommendation.
\newblock In Bonchi, F.; Domingo{-}Ferrer, J.; Baeza{-}Yates, R.; Zhou, Z.; and
  Wu, X., eds., {\em {IEEE} 16th International Conference on Data Mining,
  {ICDM} 2016, December 12-15, 2016, Barcelona, Spain},  1023--1028.
\newblock {IEEE} Computer Society.

\bibitem[\protect\citeauthoryear{Lian \bgroup et al\mbox.\egroup
  }{2018}]{LianZZCXS18}
Lian, J.; Zhou, X.; Zhang, F.; Chen, Z.; Xie, X.; and Sun, G.
\newblock 2018.
\newblock xdeepfm: Combining explicit and implicit feature interactions for
  recommender systems.
\newblock In Guo, Y., and Farooq, F., eds., {\em Proceedings of the 24th {ACM}
  {SIGKDD} International Conference on Knowledge Discovery {\&} Data Mining,
  {KDD} 2018, London, UK, August 19-23, 2018},  1754--1763.
\newblock {ACM}.

\bibitem[\protect\citeauthoryear{Liu \bgroup et al\mbox.\egroup
  }{2019}]{LiuHYCXSH19}
Liu, Q.; Huang, Z.; Yin, Y.; Chen, E.; Xiong, H.; Su, Y.; and Hu, G.
\newblock 2019.
\newblock {EKT:} exercise-aware knowledge tracing for student performance
  prediction.
\newblock {\em CoRR} abs/1906.05658.

\bibitem[\protect\citeauthoryear{Liu \bgroup et al\mbox.\egroup
  }{2020}]{LiuYCSZY20}
Liu, Y.; Yang, Y.; Chen, X.; Shen, J.; Zhang, H.; and Yu, Y.
\newblock 2020.
\newblock Improving knowledge tracing via pre-training question embeddings.
\newblock In Bessiere, C., ed., {\em Proceedings of the Twenty-Ninth
  International Joint Conference on Artificial Intelligence, {IJCAI} 2020},
  1577--1583.
\newblock ijcai.org.

\bibitem[\protect\citeauthoryear{Luo \bgroup et al\mbox.\egroup
  }{2015}]{LuoSMG15}
Luo, J.; Sorour, S.~E.; Mine, T.; and Goda, K.
\newblock 2015.
\newblock Predicting student grade based on free-style comments using word2vec
  and {ANN} by considering prediction results obtained in consecutive lessons.
\newblock In Santos, O.~C.; Boticario, J.; Romero, C.; Pechenizkiy, M.;
  Merceron, A.; Mitros, P.; Luna, J.~M.; Mihaescu, M.~C.; Moreno, P.;
  Hershkovitz, A.; Ventura, S.; and Desmarais, M.~C., eds., {\em Proceedings of
  the 8th International Conference on Educational Data Mining, {EDM} 2015,
  Madrid, Spain, June 26-29, 2015},  396--399.
\newblock International Educational Data Mining Society {(IEDMS)}.

\bibitem[\protect\citeauthoryear{Song \bgroup et al\mbox.\egroup
  }{2019}]{SongS0DX0T19}
Song, W.; Shi, C.; Xiao, Z.; Duan, Z.; Xu, Y.; Zhang, M.; and Tang, J.
\newblock 2019.
\newblock Autoint: Automatic feature interaction learning via self-attentive
  neural networks.
\newblock In Zhu, W.; Tao, D.; Cheng, X.; Cui, P.; Rundensteiner, E.~A.;
  Carmel, D.; He, Q.; and Yu, J.~X., eds., {\em Proceedings of the 28th {ACM}
  International Conference on Information and Knowledge Management, {CIKM}
  2019, Beijing, China, November 3-7, 2019},  1161--1170.
\newblock {ACM}.

\bibitem[\protect\citeauthoryear{Spector}{2018}]{Spector18}
Spector, J.~M.
\newblock 2018.
\newblock Smart learning futures: a report from the 3\({}^{\mbox{rd}}\)
  us-china smart education conference.
\newblock {\em Smart Learn. Environ.} 5(1):5.

\bibitem[\protect\citeauthoryear{Su \bgroup et al\mbox.\egroup
  }{2018}]{SuLLHYCDWH18}
Su, Y.; Liu, Q.; Liu, Q.; Huang, Z.; Yin, Y.; Chen, E.; Ding, C. H.~Q.; Wei,
  S.; and Hu, G.
\newblock 2018.
\newblock Exercise-enhanced sequential modeling for student performance
  prediction.
\newblock In McIlraith, S.~A., and Weinberger, K.~Q., eds., {\em Proceedings of
  the Thirty-Second {AAAI} Conference on Artificial Intelligence, (AAAI-18),
  the 30th innovative Applications of Artificial Intelligence (IAAI-18), and
  the 8th {AAAI} Symposium on Educational Advances in Artificial Intelligence
  (EAAI-18), New Orleans, Louisiana, USA, February 2-7, 2018},  2435--2443.
\newblock {AAAI} Press.

\bibitem[\protect\citeauthoryear{Sun \bgroup et al\mbox.\egroup
  }{2011}]{SunHYYW11}
Sun, Y.; Han, J.; Yan, X.; Yu, P.~S.; and Wu, T.
\newblock 2011.
\newblock Pathsim: Meta path-based top-k similarity search in heterogeneous
  information networks.
\newblock {\em Proc. {VLDB} Endow.} 4(11):992--1003.

\bibitem[\protect\citeauthoryear{Sweeney \bgroup et al\mbox.\egroup
  }{2016}]{SweeneyLRJ16}
Sweeney, M.; Lester, J.; Rangwala, H.; and Johri, A.
\newblock 2016.
\newblock Next-term student performance prediction: {A} recommender systems
  approach.
\newblock In Barnes, T.; Chi, M.; and Feng, M., eds., {\em Proceedings of the
  9th International Conference on Educational Data Mining, {EDM} 2016, Raleigh,
  North Carolina, USA, June 29 - July 2, 2016}, ~7.
\newblock International Educational Data Mining Society {(IEDMS)}.

\bibitem[\protect\citeauthoryear{Tan, Xiang, and Zhou}{2015}]{TanXZ15}
Tan, M.; Xiang, B.; and Zhou, B.
\newblock 2015.
\newblock Lstm-based deep learning models for non-factoid answer selection.
\newblock {\em CoRR} abs/1511.04108.

\bibitem[\protect\citeauthoryear{Thai{-}Nghe \bgroup et al\mbox.\egroup
  }{2010}]{Thai-NgheDKS10}
Thai{-}Nghe, N.; Drumond, L.; Krohn{-}Grimberghe, A.; and Schmidt{-}Thieme, L.
\newblock 2010.
\newblock Recommender system for predicting student performance.
\newblock In Manouselis, N.; Drachsler, H.; Verbert, K.; and Santos, O.~C.,
  eds., {\em Proceedings of the 1st Workshop on Recommender Systems for
  Technology Enhanced Learning, RecSysTEL 2010, Barcelona, Spain, September
  29-30, 2010}, volume~1 of {\em Procedia Computer Science},  2811--2819.
\newblock Elsevier.

\bibitem[\protect\citeauthoryear{Vaswani \bgroup et al\mbox.\egroup
  }{2017}]{VaswaniSPUJGKP17}
Vaswani, A.; Shazeer, N.; Parmar, N.; Uszkoreit, J.; Jones, L.; Gomez, A.~N.;
  Kaiser, L.; and Polosukhin, I.
\newblock 2017.
\newblock Attention is all you need.
\newblock In Guyon, I.; von Luxburg, U.; Bengio, S.; Wallach, H.~M.; Fergus,
  R.; Vishwanathan, S. V.~N.; and Garnett, R., eds., {\em Advances in Neural
  Information Processing Systems 30: Annual Conference on Neural Information
  Processing Systems 2017, 4-9 December 2017, Long Beach, CA, {USA}},
  5998--6008.

\bibitem[\protect\citeauthoryear{Vie and Kashima}{2019}]{VieK19}
Vie, J., and Kashima, H.
\newblock 2019.
\newblock Knowledge tracing machines: Factorization machines for knowledge
  tracing.
\newblock In {\em The Thirty-Third {AAAI} Conference on Artificial
  Intelligence, {AAAI} 2019, The Thirty-First Innovative Applications of
  Artificial Intelligence Conference, {IAAI} 2019, The Ninth {AAAI} Symposium
  on Educational Advances in Artificial Intelligence, {EAAI} 2019, Honolulu,
  Hawaii, USA, January 27 - February 1, 2019},  750--757.
\newblock {AAAI} Press.

\bibitem[\protect\citeauthoryear{Wang \bgroup et al\mbox.\egroup
  }{2019}]{WangWX00C19}
Wang, X.; Wang, D.; Xu, C.; He, X.; Cao, Y.; and Chua, T.
\newblock 2019.
\newblock Explainable reasoning over knowledge graphs for recommendation.
\newblock In {\em The Thirty-Third {AAAI} Conference on Artificial
  Intelligence, {AAAI} 2019, The Thirty-First Innovative Applications of
  Artificial Intelligence Conference, {IAAI} 2019, The Ninth {AAAI} Symposium
  on Educational Advances in Artificial Intelligence, {EAAI} 2019, Honolulu,
  Hawaii, USA, January 27 - February 1, 2019},  5329--5336.
\newblock {AAAI} Press.

\bibitem[\protect\citeauthoryear{Xian \bgroup et al\mbox.\egroup
  }{2019}]{XianFMMZ19}
Xian, Y.; Fu, Z.; Muthukrishnan, S.; de~Melo, G.; and Zhang, Y.
\newblock 2019.
\newblock Reinforcement knowledge graph reasoning for explainable
  recommendation.
\newblock In Piwowarski, B.; Chevalier, M.; Gaussier, {\'{E}}.; Maarek, Y.;
  Nie, J.; and Scholer, F., eds., {\em Proceedings of the 42nd International
  {ACM} {SIGIR} Conference on Research and Development in Information
  Retrieval, {SIGIR} 2019, Paris, France, July 21-25, 2019},  285--294.
\newblock {ACM}.

\bibitem[\protect\citeauthoryear{Xian \bgroup et al\mbox.\egroup
  }{2020}]{XianFZGCHG0MMZ20}
Xian, Y.; Fu, Z.; Zhao, H.; Ge, Y.; Chen, X.; Huang, Q.; Geng, S.; Qin, Z.;
  de~Melo, G.; Muthukrishnan, S.; and Zhang, Y.
\newblock 2020.
\newblock {CAFE:} coarse-to-fine neural symbolic reasoning for explainable
  recommendation.
\newblock In d'Aquin, M.; Dietze, S.; Hauff, C.; Curry, E.; and
  Cudr{\'{e}}{-}Mauroux, P., eds., {\em {CIKM} '20: The 29th {ACM}
  International Conference on Information and Knowledge Management, Virtual
  Event, Ireland, October 19-23, 2020},  1645--1654.
\newblock {ACM}.

\bibitem[\protect\citeauthoryear{Xing \bgroup et al\mbox.\egroup
  }{2015}]{XingGPG15}
Xing, W.; Guo, R.; Petakovic, E.; and Goggins, S.~P.
\newblock 2015.
\newblock Participation-based student final performance prediction model
  through interpretable genetic programming: Integrating learning analytics,
  educational data mining and theory.
\newblock {\em Comput. Hum. Behav.} 47:168--181.

\bibitem[\protect\citeauthoryear{Yao \bgroup et al\mbox.\egroup
  }{2019}]{YaoLCWZ19}
Yao, H.; Lian, D.; Cao, Y.; Wu, Y.; and Zhou, T.
\newblock 2019.
\newblock Predicting academic performance for college students: {A} campus
  behavior perspective.
\newblock {\em {ACM} Trans. Intell. Syst. Technol.} 10(3):24:1--24:21.

\bibitem[\protect\citeauthoryear{Yeung}{2019}]{Yeung19}
Yeung, C.
\newblock 2019.
\newblock Deep-irt: Make deep learning based knowledge tracing explainable
  using item response theory.
\newblock In Desmarais, M.~C.; Lynch, C.~F.; Merceron, A.; and Nkambou, R.,
  eds., {\em Proceedings of the 12th International Conference on Educational
  Data Mining, {EDM} 2019, Montr{\'{e}}al, Canada, July 2-5, 2019}.
\newblock International Educational Data Mining Society {(IEDMS)}.

\bibitem[\protect\citeauthoryear{Zhang \bgroup et al\mbox.\egroup
  }{2017}]{ZhangSPST17}
Zhang, X.; Sun, G.; Pan, Y.; Sun, H.; and Tan, J.
\newblock 2017.
\newblock Poor performance discovery of college students based on behavior
  pattern.
\newblock In {\em 2017 {IEEE} SmartWorld, Ubiquitous Intelligence {\&}
  Computing, Advanced {\&} Trusted Computed, Scalable Computing {\&}
  Communications, Cloud {\&} Big Data Computing, Internet of People and Smart
  City Innovation, SmartWorld/SCALCOM/UIC/ATC/CBDCom/IOP/SCI 2017, San
  Francisco, CA, USA, August 4-8, 2017},  1--8.
\newblock {IEEE}.

\bibitem[\protect\citeauthoryear{Zhou \bgroup et al\mbox.\egroup
  }{2018}]{ZhouHHZT18}
Zhou, Y.; Huang, C.; Hu, Q.; Zhu, J.; and Tang, Y.
\newblock 2018.
\newblock Personalized learning full-path recommendation model based on {LSTM}
  neural networks.
\newblock {\em Inf. Sci.} 444:135--152.

\end{thebibliography}
\end{document}